\documentclass[
superscriptaddress,
reprint,
amsmath,amssymb,
aps,
prl,
floatfix,
longbibliography,
twocolumn
]{revtex4-2}

\usepackage{mathrsfs}
\usepackage{graphicx}
\usepackage{dcolumn}
\usepackage{bm}
\usepackage{chemformula}
\usepackage{braket}
\usepackage{multirow}
\usepackage{booktabs}  
\usepackage{tabularx}  
\usepackage{adjustbox}  
\usepackage{hyperref}

\hypersetup{
    colorlinks=true,
    linkcolor=blue,
    citecolor=blue,
    urlcolor=blue,
}

\usepackage{xr}
\begin{document}

\title{General Theory for Ferroelectric Control of Spin Splitting in Collinear Antiferromagnets}

\author{Zhihao Dai\textsuperscript{*}}
\affiliation{Key Laboratory of Computational Physical Sciences (Ministry of Education), Institute of Computational Physical Sciences, State Key Laboratory of Surface Physics and Department of Physics, Fudan University, Shanghai 200433, China}

\author{Yingwei Chen\textsuperscript{*}}
\affiliation{Key Laboratory of Computational Physical Sciences (Ministry of Education), Institute of Computational Physical Sciences, State Key Laboratory of Surface Physics and Department of Physics, Fudan University, Shanghai 200433, China}

\author{Junyi Ji}
\affiliation{Beijing National Laboratory for Condensed Matter Physics and Institute of Physics, Chinese Academy of Sciences, Beijing 100190, China}

\author{Chaoyu He\textsuperscript{$\dagger$}}
\affiliation{School of Physics and Optoelectronics, Xiangtan University, Xiangtan 411105, China}
\affiliation{Center for Quantum Science and Technology, Shanghai University, Shanghai 200444, China}

\author{Hongjun Xiang\textsuperscript{$\ddagger$}}
\affiliation{Key Laboratory of Computational Physical Sciences (Ministry of Education), Institute of Computational Physical Sciences, State Key Laboratory of Surface Physics and Department of Physics, Fudan University, Shanghai 200433, China}

\begin{abstract}
Electrical control of magnetism is crucial for next-generation spintronics. 
While recent advances have demonstrated ferroelectric switching in two-dimensional magnets, a general design strategy spanning different dimensionalities remains elusive. 
Here, we develop a group-theoretical framework for achieving ferroelectric control of spin splitting in collinear antiferromagnets, including altermagnets and compensated ferrimagnets. 
By systematically classifying switching operators through symmetry analysis, we identify a universal pathway for the simultaneous reversal of electric polarization and nonrelativistic spin splitting.
We validate this approach in three representative systems: quasi-one-dimensional $(6,14)$ Zigzag graphene nanoribbons, two-dimensional~\ch{Nb3I8}, and three-dimensional altermagnetic~\ch{MnSe2}. 
Our work establishes a versatile design paradigm for magnetoelectric devices and expands the functional landscape of low-power spintronic materials beyond the low-dimensional limit.
\end{abstract}
\date{\today}
\maketitle

\textsuperscript{*}These authors contributed equally to this work.\\
\textsuperscript{$\dagger$}Corresponding author: \texttt{hechaoyu@xtu.edu.cn}\\
\textsuperscript{$\ddagger$}Corresponding author: \texttt{hxiang@fudan.edu.cn}

\begin{figure}[t] 
    \centering
    \includegraphics[width=1\linewidth]{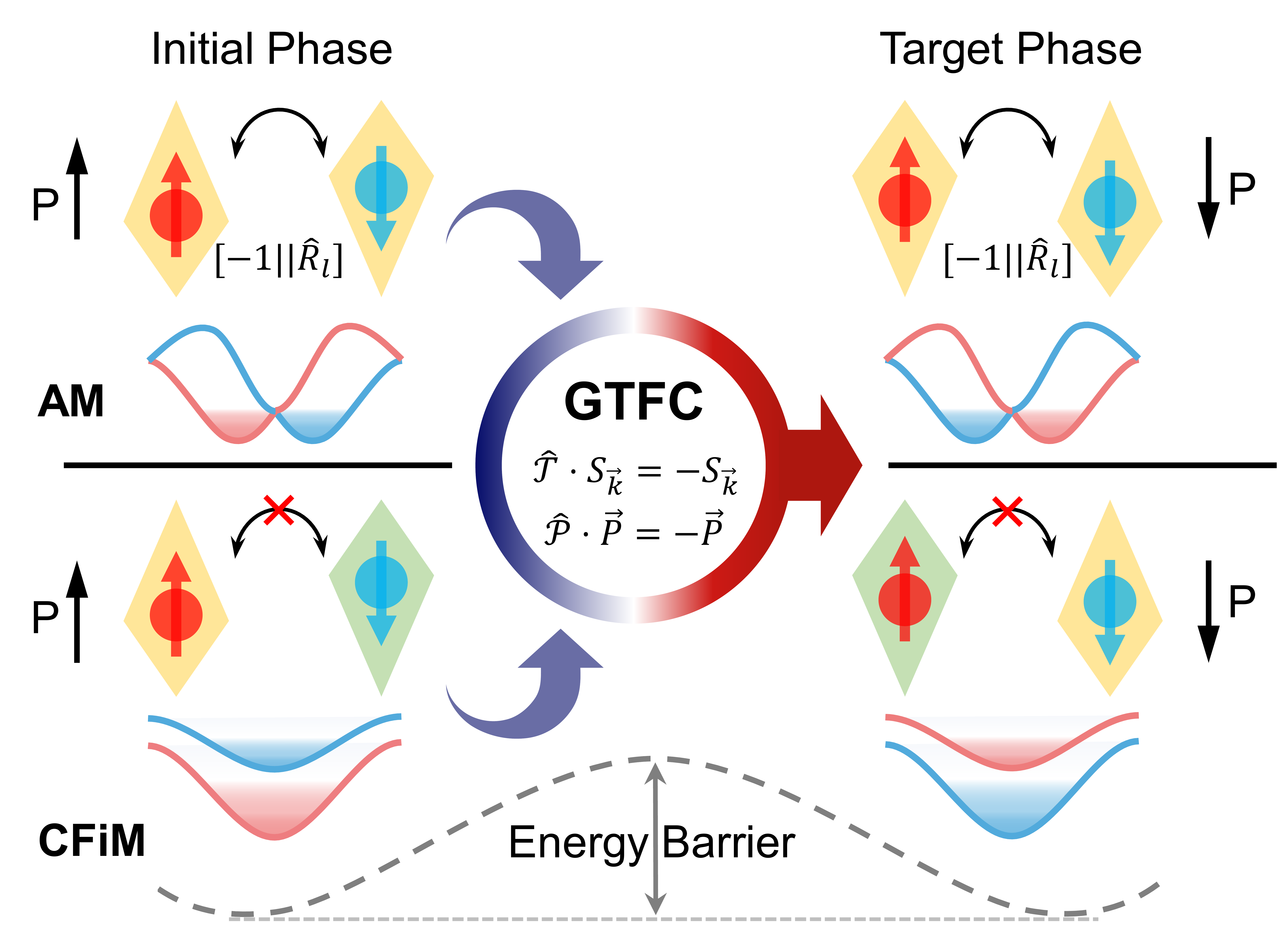} 
    \caption{Schematic diagrams of the general theory for ferroelectric control (GTFC) of spin splitting in (a) altermagnets (AMs) and (b) collinear compensated ferrimagnets (CFiMs). 
    Red and blue arrows (lines) denote opposite magnetic moments (spin channels). 
    Black arrows indicate electric polarization $\vec P$.
    Yellow and green shaded regions represent distinct ligand environments around magnetic sites.
    In AMs, these environments are connected by an intrinsic symmetry $\hat R_l$, while in CFiMs no such symmetry exists.    }
    \label{Fig1} 
\end{figure}

\textit{Introduction}——Antiferromagnets (AFMs) offer compelling advantages for spintronic applications, including ultrafast dynamics, robustness against external magnetic fields, and negligible stray fields~\cite{rev_2018_NatPhy_Jungwirth}.
However, their net zero magnetization makes magnetic signals difficult to detect and manipulate.
Recently, altermagnets (AMs)~\cite{2004_Wu_PRL, 2007_Wu_PRB, 2019_Hayami_JPSJ, 2020_PRB_Yuan, 2020_Hayami_PRB, 2021_NC_Ma, 2022_PRX_Smejkal_2, 2025_PRL_Gu, 2025_PRL_Duan} have emerged as materials that retain a net zero magnetization while generating nonrelativistic spin splitting (NRSS), thereby overcoming this limitation.
This unique property gives rise to diverse spintronic phenomena, such as spin-polarized currents~\cite{2021_PRL_Gonzalez}, the anomalous Hall effect~\cite{2023_PRB_Nguyen,2023_PRL_Gonzalez,2023_NPJCM_Guo,2025_Nat_Zhou}, chiral magnon excitations~\cite{2022_PRX_Smejkal,2024_AdvFM_Bai}, and magneto-optical responses~\cite{2022_PRX_Smejkal,2024_AdvFM_Bai}.
Unlike AMs, where magnetic sublattices are connected by specific lattice symmetry~$\hat R_l$, such as rotational $\hat{C}_n$ or mirror $\hat M$, the collinear compensated ferrimagnets (CFiMs)~\cite{2020_APL_Finley,2022_NatMat_Kim,2024_PRL_Yuan,2025_PRL_Liu} achieve zero net magnetization through appropriate band filling, yet still host NRSS in momentum space [see Fig.~\ref{Fig1}].
Both theoretical and experimental studies have demonstrated the realization of CFiMs in diverse material platforms, notably via elemental substitution~\cite{1995_PRL_Van,2017_PRB_Stinshoff,2025_PRL_Liu,2022_SciRep_Semboshi} and in two-dimensional van der Waals materials via strategies such as Janus structures~\cite{2006_Nat_Son,2017_NatNano_Lu,2025_PRL_Liu} and staggered potentials~\cite{2025_PRL_Liu}.

To realize energy-efficient devices based on unconventional magnetism, electrical reversal of their NRSS is pivotal~\cite{ 2016_Sci_Wadley, rev_2022_Nat_Yang, 2024_SciAdv_Han, 2025_PRL_Chen}.
Despite progress in methods like spin-orbit torque-driven switching~\cite{2024_SciAdv_Han, 2025_PRL_Chen}, their dependence on charge current injection fundamentally limits efficiency and leads to significant energy dissipation~\cite{2025_PRL_Zhu}.
In this context, multiferroic materials~\cite{1994_Ferro_Schmid}, which exhibit coupled ferroelectric and magnetic orders, offer a promising low‑power alternative by enabling control of spin polarization through ferroelectric polarization switching~\cite{2025_AdvM_Sun, 2025_PRL_Gu, 2025_PRL_Zhu, 2025_PRB_Cheng, 2025_PRB_Mavani,2023_npjCM_Liu, 2025_NL_Chen}. 
Indeed, several recent studies have shown that ferroelectric reversal can efficiently switch NRSS~\cite{2025_PRL_Zhu, 2025_PRB_Cheng, 2025_PRB_Mavani, 2025_AdvM_Sun, 2023_npjCM_Liu}. 
Despite this progress, most works to date have been confined to 2D materials and out-of-plane polarization, significantly restricting the material search space.
To overcome this limitation and unlock ferroelectric switchable NRSS across a broader spectrum of materials, a general design strategy, which transcends specific dimensionalities and polarization orientations, is urgently needed.

In this Letter, we establish a general theory for ferroelectric control (GTFC) of spin splitting in collinear AFM, encompassing both AMs and CFiMs as shown in Fig.~\ref{Fig1}.
Our approach systematically identifies feasible switching operators across different dimensionalities and polarization orientations, which enables the prediction of a symmetry-related multiferroic state and the establishment of a switching pathway for synchronous ferroelectric and NRSS reversal.
To illustrate its universality, first-principles calculations are performed on quasi-one-dimensional (1D) $(6,14)$ zigzag graphene nanoribbons, two-dimensional (2D)~\ch{Nb3I8}, and three-dimensional (3D)~\ch{MnSe2}. 
These case studies exemplify how our symmetry-based framework guides the discovery of reversible switching pathways between multiferroic states, a crucial step towards realizing low-energy spintronic devices.

\textit{Group Theory Analysis of Ferroelectric Control in Multiferroic States}——In a multiferroic state \( |M\rangle \), the reversal of NRSS by flipping the ferroelectric polarization \( \vec{P} \) corresponds to a symmetry operator.
Here, \( S_{\vec{k}} \equiv \varepsilon_{\vec{k}}^{\uparrow} - \varepsilon_{\vec{k}}^{\downarrow} \) denotes the spin splitting between opposite spin channels along certain paths in the Brillouin zone.
Since spin-orbit coupling is neglected, this switching mechanism must be analyzed using spin-group theory, which unifies spin and spatial symmetries.
In a common operation, such an operator is denoted as $[g_s||g_l|\tau]$, combining a spin-space operation $g_s$, a real-space point-group operation $g_l$, and a translation $\tau$~\cite{2022_PRX_Liu}.
A spin point group contains only point symmetries $[g_s||g_l]$.
Currently, research~\cite{2025_PRL_Zhu, 2025_PRB_Mavani, 2025_AdvM_Sun, 2023_npjCM_Liu, 2025_PRB_Cheng, 2025_NL_Wang} on the ferroelectric control of collinear magnetism predominantly focuses on 2D AMs.
To systematically analyze how dimensionality and intrinsic symmetries jointly govern the ferroelectric switching of NRSS in unconventional magnetism, we develop a general group-theoretical framework.

Consider the two switchable multiferroic states as $\ket{M_1}$ and $\ket{M_2}$, characterized by two order parameters $(S_{\vec{k}},\vec{P})$ and $(-S_{\vec{k}},-\vec{P})$, respectively. 
A trivial operator to switch these states is the composite action of spatial inversion $\hat{\mathcal{P}}$ and time reversal $\hat{\mathcal{T}}$:
\begin{equation}
\hat{\mathcal{P}}\hat{\mathcal{T}}\ket{M_1}=\ket{M_2}
\end{equation}
where $\hat{\mathcal{T}} S_{\vec{k}} =\hat{\mathcal{T}} \left(\varepsilon_{\vec{k}}^{\uparrow} - \varepsilon_{\vec{k}}^{\downarrow}\right)= -S_{-\vec{k}}=-S_{\vec{k}}$ and $\hat{\mathcal{P}}\vec{P} = -\vec{P}$. 
The even parity of $S_{\vec{k}}$ is ensured by the collinear magnetic moments.
To exhaustively enumerate all possible operators capable of switching these states, we first determine the multiferroic stabilizer subgroup $\{\hat{R}\}$, which maintains $\vec S_{\vec k}$ and $\vec P$ invariance, ensuring that all operators in $\{\hat{R}\}$ do not switch multiferroic states:
\begin{equation}
\forall \  \hat r \in \{\hat{R}\} : \hat r \ket{M_i}= \ket{M_i}
\end{equation}
The complete set of switching operators $\{\hat{T}\}$ is then obtained by constructing the left coset of $\{\hat{R}\}$ with respect to the combined $\mathcal{\hat{T}\hat{P}}$ operator:
\begin{equation}
\{\hat{T}\} \equiv \{\hat t \ | \ \hat t =\mathcal{\hat{T}\hat{P}} \cdot\ \hat r, \ \hat r \in \{\hat{R}\} \}
\end{equation}
Every operator in $\{\hat{T}\}$ simultaneously reverses both order parameters, mapping $\ket{M_1}$ to $\ket{M_2}$.
Furthermore, by disregarding the intrinsic symmetries of the material, the minimal multiferroic stabilizer subgroup $\{\hat{R}_{\mathcal{D},i}\}$ is obtained, which depends only on the dimension $\mathcal{D}$ of momentum space and the polarization direction $i$. 
The corresponding switching set $\{\hat{T}_{\mathcal{D},i}\}$ is then defined as the minimal set of symmetry-allowed switching operators, as shown in Table~\ref{Table1}.
This minimal set is universal for systems with low intrinsic symmetry, for example, in certain CFiMs where no intrinsic symmetries are present.
The explicit derivation is provided in the Supplemental Material Sec.~I~\cite{SM_file}.
In materials with higher intrinsic crystallographic symmetry, such as AMs, the full multiferroic stabilizer group $\{\hat{R}\}$ expands beyond this minimal form due to a nontrivial intrinsic symmetry subgroup, thereby broadening the set of available switching operators.
This expansion provides richer pathways for ferroelectric control, especially in 3D materials.
Overall, the intrinsic symmetries broaden the landscape of available switching operators, yet these additional pathways are contingent on specific crystallographic symmetries and may be compromised by symmetry-lowering perturbations. 
In contrast, the minimal switching set $\{\hat{T}_{\mathcal{D},i}\}$, determined solely by dimensionality and polarization direction, offers fewer but more robust pathways that remain valid regardless of material-specific symmetry details.
Our general theoretical framework is compatible with existing studies~\cite{2025_PRL_Zhu, 2025_PRB_Mavani, 2025_AdvM_Sun, 2023_npjCM_Liu, 2025_NL_Wang, 2025_NL_Wang_ding, 2025_PRB_Cheng, 2025_PRL_Gu} on the coupling between ferroelectricity and NRSS in unconventional magnetism, as detailed in the Supplemental Material, Table~S2~\cite{SM_file}.
To further validate and illustrate its predictive power, first-principles calculations based on density functional theory (DFT)~\cite{1996_CMS_Kresse, 1999_PRB_Kresse, 1996_PRL_Perdew, 2010_JCP_Grimme, 1998_PRB_Dudarev, 1994_RMP_Resta,1992_Fe_Resta, 1993_PRB_King-Smith, 2000_JCP_Henkelman, 2015_PRB_Ma} are performed, presenting concrete examples of ferroelectric-controlled NRSS switching across different dimensionalities in the following sections.

\begin{table}[h]
\caption{Enumeration of switching operator sets $\{\hat{T}\}$ for unconventional magnetism across dimensionalities. 
For low-dimensional systems, $P_\parallel$ ($P_\perp$) denotes polarization parallel (perpendicular) to the periodic direction. 
In 1D, $\perp_1$ and $\perp_2$ distinguish the two orthogonal axes perpendicular to the periodic direction, with $\perp_1$ aligned with the polarization direction for the $P_\perp$ case. }
\label{Table1}
\begin{adjustbox}{width=8.5cm, center}
\renewcommand{\arraystretch}{1.5}
\begin{tabular}{w{c}{1.5cm}|w{c}{3.5cm}|w{c}{3.5cm}} 
\noalign{\hrule height 0.5pt}
\noalign{\vskip 1pt} 
\noalign{\hrule height 0.5pt}

$\{\hat{T}_{\mathcal{D},i}\}$ & $P_\parallel$ & $P_\perp$   \\ 
\noalign{\hrule height 0.75pt}
\multirow{2}{*}{1D} & $[-1||\hat{\mathcal{P}}]$, $[-1|| \hat{\mathcal{P}}\hat C_n]$ & $[-1||\hat{\mathcal{P}}]$, $[-1||\hat M_{\perp_1}]$  \\
                    & $[-1||\hat2_{\perp}\hat C_n]$  & $[-1||\hat2_{\parallel}]$, $[-1||\hat2_{\perp_2}]$  \\
\noalign{\hrule height 0.45pt}
2D                  & $[-1||\hat{\mathcal{P}}]$, $[-1||\hat 2_{\perp}]$ & $[-1||\hat{\mathcal{P}}]$, $[-1||\hat M_{\perp}]$ \\
\noalign{\hrule height 0.45pt}
3D                  & \multicolumn{2}{c}{$[-1||\hat{\mathcal{P}}]$}  \\
\noalign{\hrule height 0.5pt}
\noalign{\vskip 1pt} 
\noalign{\hrule height 0.5pt}
\end{tabular}
\end{adjustbox}
\end{table}

%
\textit{Quasi-1D sliding ferroelectric control in (6,14)-ZGNRs}——Zigzag graphene nanoribbons (ZGNRs) constitute a carbon-based spintronic platform featuring magnetically ordered edge states stabilized by exchange interactions at the zigzag edges~\cite{2006_Nat_Son, 2008_PRL_Yazyev, 2014_NN_Han}. 
Experimentally, such states have been realized via nitrogen-passivated edges~\cite{2021_Nat_Blackwell} and atomically sharp Mo-graphene interfaces~\cite{2026_NT_Wei}. 
Here, we demonstrate a novel approach to control bilayer ZGNRs with mismatched zigzag widths, utilizing sliding ferroelectricity to achieve synchronous switching of both NRSS and electrical polarization.

\begin{figure}[h] 
    \centering
    \includegraphics[width=1\linewidth]{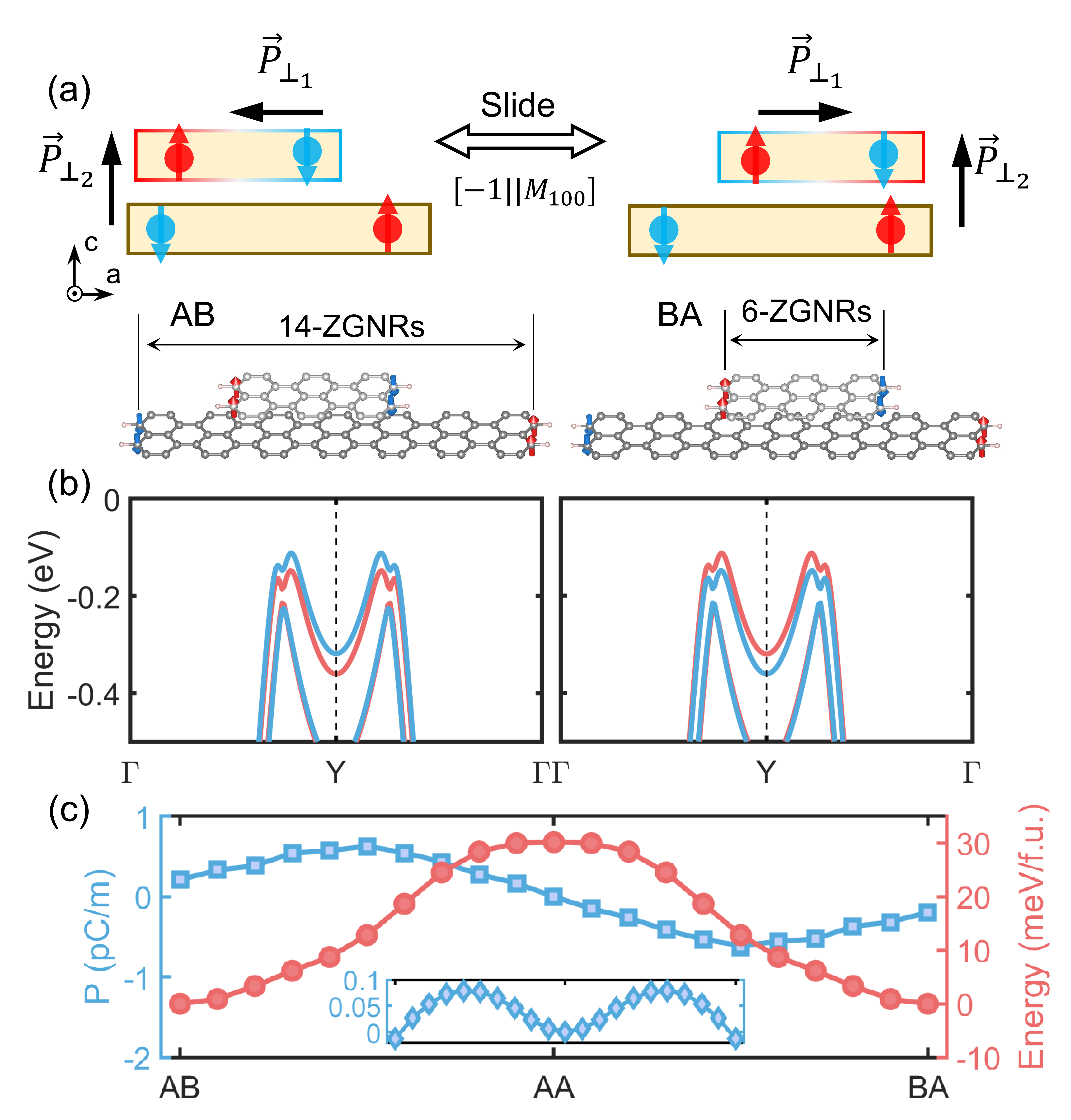} 
    \caption{Sliding ferroelectric switching in quasi‑1D (6,14)-ZGNRs.
    (a) Schematic of the multiferroic switching mediated by the $[-1||\hat M_{100}]$ operator. 
    Red (blue) arrows denote spin‑up (spin‑down) magnetic moments at the edges. 
    The blue-to-red gradient indicates the sliding direction from the AA stacking; black arrows show the electric polarization.
    (b) Band structures for the AB (left) and BA (right) stackings. Red and blue correspond to opposite spin channels.
    (c) Evolution along the sliding path from AB to BA: energy barrier (red) and in‑plane (squares)/out‑of‑plane (diamonds) polarization (blue).
   }
    \label{Fig2} 
\end{figure}

As shown in Fig.~\ref{Fig2}(a), the number of zigzag chains in the top and bottom layers is denoted by $(n,m)$-ZGNRs.
The $(6,14)$-ZGNRs, in which the interlayer AFM coupling is energetically favorable over the FM configuration by $0.43 \ \mathrm{meV}$ per cell, exhibit a lower sliding barrier. 
The AA stacking configuration, which is a higher-energy, nonpolar conventional AFM, hosts degenerate spin bands protected by the combined symmetry \([-1||\hat{M}_{100}]\). 
Sliding the top layer along the negative direction of the a-axis from the AA stacking breaks the $[-1||\hat{M}_{100}]$ symmetry, driving the system into a CFiM state, labeled AB stacking, that exhibits NRSS. 
This sliding explicitly breaks spatial inversion, endowing the system with an electric polarization and thus establishing a multiferroic state.
According to the classification in Table~\ref{Table1}, applying the switching operator $[-1||\hat{M}_{100}]$ to this AB state generates a symmetry-related partner state with simultaneously reversed in-plane polarization and NRSS.
From the sliding perspective, this partner state corresponds to BA stacking, which can equivalently be reached by sliding the top layer from the AA configuration by the same distance in the opposite direction. 
These two degenerate CFiM ground states, AB and BA stacking, are energetically favorable over the AA phase by $30.16\ \mathrm{meV/f.u.}$.

The calculated edge magnetic moment is $0.137\pm0.001 \ \mu_B$. 
The AB and BA stackings exhibit a maximum spin splitting of \(\mp 42.0\) meV at the valence band maximum (VBM) and out-of-plane polarizations of \(\pm 0.210\) pC/m, respectively, while both share a common in-plane polarization of \(-0.013\) pC/m, as shown in Fig.~\ref{Fig2}(b).
The evolution of polarization and the energy barrier along the sliding path are plotted in Fig.~\ref{Fig2}(c).
Notably, the in-plane polarization displays odd parity while the out-of-plane component shows even parity as a function of the sliding coordinate, in full agreement with the symmetry-based predictions in Table~\ref{Table1}.
These findings indicate that an applied in-plane electric field may provide a direct means to manipulate the NRSS via the interlayer sliding mechanism.

\textit{2D sliding ferroelectric control in~\ch{Nb3I8}}——\ch{Nb3I8} is a van der Waals layered transition-metal halide featuring a breathing-kagome lattice, which has been experimentally realized in bulk and few-layer forms~\cite{2019_RRL_Kim,2022_CM_Regmi}.
Its monolayer exhibits room-temperature ferromagnetism with a Curie temperature about 307 K~\cite{2020_PRR_Conte}, while its bilayer shows stacking-dependent interlayer antiferromagnetism~\cite{2020_PRR_Conte,2025_arXiv_Zhang} coupled with sliding ferroelectricity, enabling tunable multiple-state polarized AFM~\cite{2025_PRB_Xie}. 
These properties, supported by both theoretical predictions and experimental characterizations, make~\ch{Nb3I8} an exemplary platform for exploring electrically controlled spin polarization switching and multistate spintronic memory devices.

\begin{figure}[h] 
    \centering
    \includegraphics[width=1\linewidth]{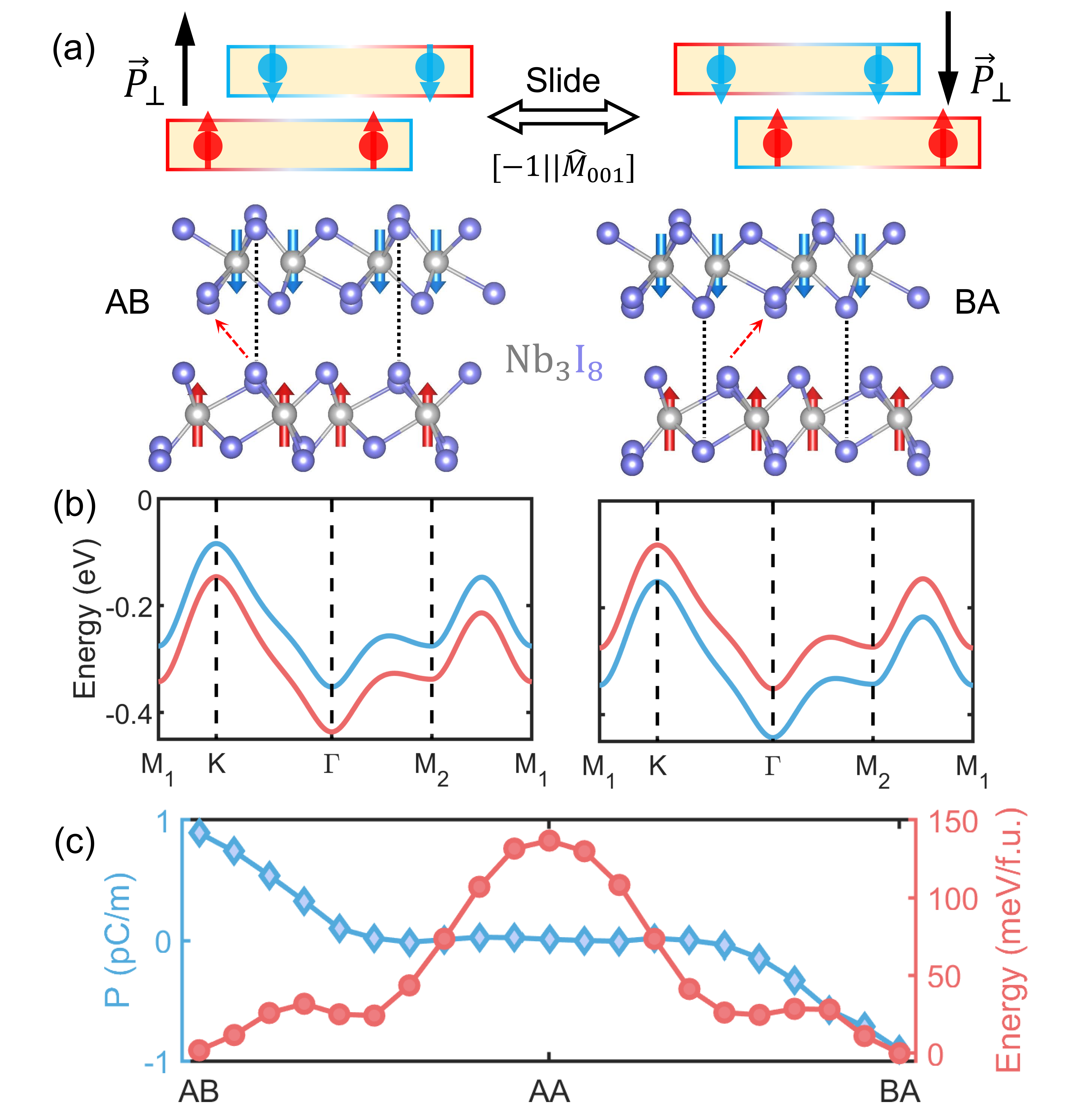} 
    \caption{Sliding ferroelectric switching in bilayer~\ch{Nb3I8}. 
    (a) Schematic of the multiferroic switching mechanism. 
    Switching is driven by the interlayer sliding operator $[-1||\hat M_{001}]$.
    Red (blue) arrows denote spin‑up (spin‑down) moments; black arrows indicate the out-of-plane polarization.
    (b) Band structures for AB (left) and BA (right) stackings. 
    Red and blue correspond to opposite spin channels.
    (c) Evolution along the sliding path from AB to BA: energy barrier (red) and polarization (blue).
    }
    \label{Fig3} 
\end{figure}

A bilayer \ch{Nb3I8} system~\cite{2021_C2DB} is constructed in the head-to-head antiferroelectric configuration, as shown in Fig.~\ref{Fig3}(a), which exhibits stronger polarization than its tail-to-tail counterpart~\cite{2025_PRB_Xie}.
Compared to (6,14)-ZGNRs, the Kramers degeneracy in AA-stacked bilayer \ch{Nb3I8} is preserved by the symmetry $[-1||\hat M_{001}]$.
Similarly, in bilayer \ch{Nb3I8}, interlayer sliding along the \((-1/3, 1/3)\) or \((1/3, -1/3)\) direction breaks this symmetry present in the AA stacking.
This symmetry breaking drives the system into CFiM states exhibiting stable NRSS and induces a finite out-of-plane electric polarization, thereby establishing multiferroic phases. 
As classified in Table~\ref{Table1}, the resulting AB and BA stackings are connected by the \([-1||\hat{M}_{001}]\) operator, which guarantees the synchronous reversal of both the out-of-plane polarization and the NRSS between these two degenerate multiferroic states.
Our first-principles calculations yield a magnetic moment of $0.265\pm 0.002 \ \mathrm{\mu_B}$ per~\ch{Nb} ion. 
As shown in Fig.~\ref{Fig3}(b), the VBM at $\Gamma$ exhibits a maximum spin splitting of $91.5 \ \mathrm{meV}$. 
The out-of-plane polarization is $\pm0.900 \ \mathrm{pC/m}$ for the AB (BA) state. 
The energy barrier for switching is calculated to be $136.73 \ \mathrm{meV/f.u.}$. 
The evolution of polarization along the sliding path, shown in Fig.~\ref{Fig3}(c), and the identified switching operator are fully consistent with our symmetry analysis presented in Table~\ref{Table1}.
This reversible out-of-plane polarization directly implies that the NRSS can likewise be controlled by an out-of-plane electric field.

\begin{figure}[t] 
    \centering
    \includegraphics[width=1\linewidth]{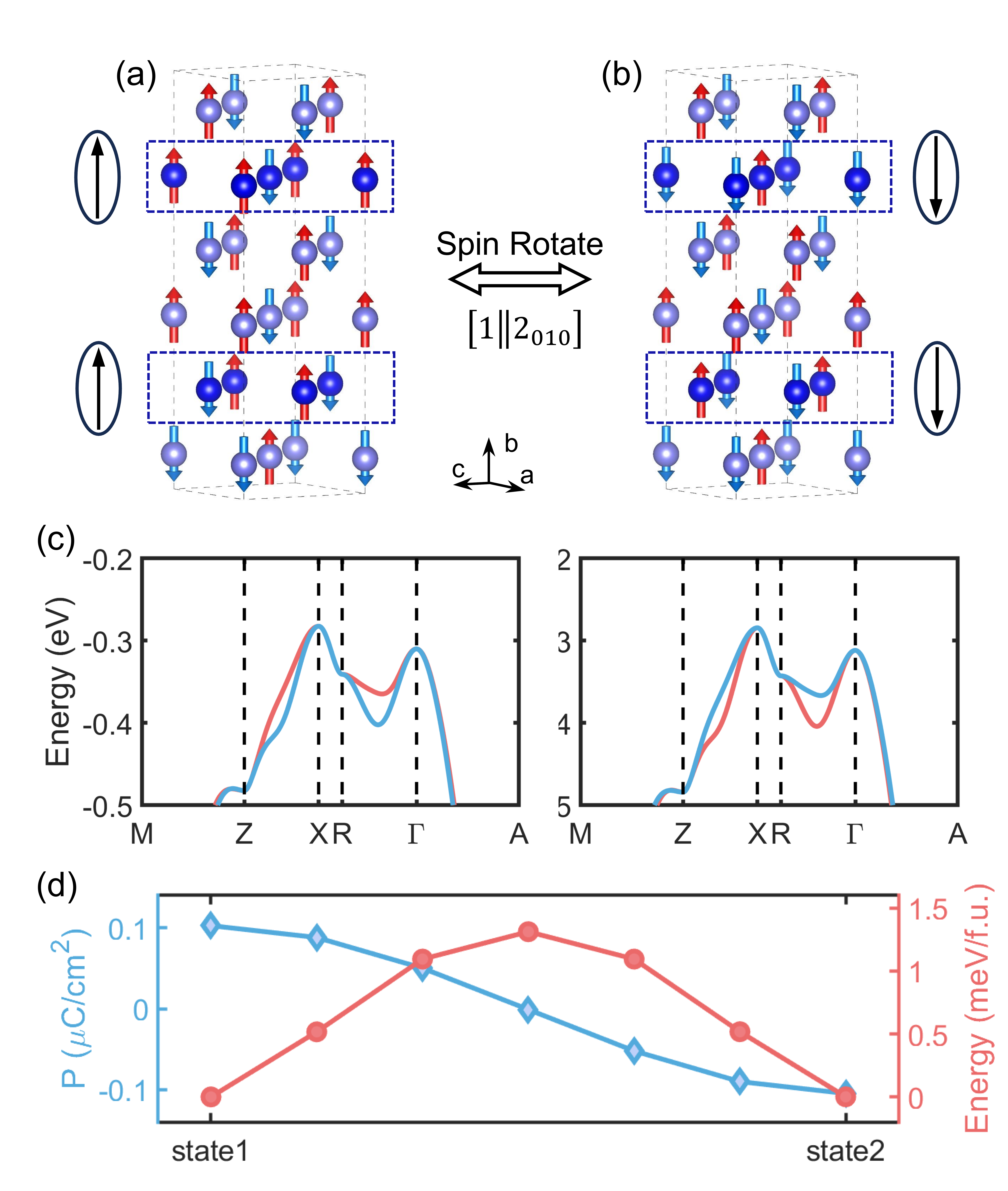} 
    \caption{Multiferroic switching in supercell altermagnets~\ch{MnSe2}.
    Spin configurations of state 1 (a) and state 2 (b). 
    Switching is mediated by the operator $[1||\hat 2_{010}]$, which reverses the moments on the second and fifth Mn layers.
    Red (blue) arrows denote spin‑up (spin‑down) moments; arrows inside ellipses indicate the N\'{e}el vector direction.
    (c) Band structures for state 1 (left) and state 2 (right). 
    Red and blue correspond to opposite spin channels.
    (d) Evolution of the system energy (red) and electric polarization (blue) during the moment‑reversal process.
    }
    \label{Fig4} 
\end{figure}

\textit{3D ferroelectric control in~\ch{MnSe2}}——\ch{MnSe2}~\cite{2016_MAGNDATA, 1987_SSC_Chattopadhyay} is a representative supercell altermagnet~\cite{2024_AdvFM_Bai, 2024_PRB_Jaeschke, 2024_ACS_Wei} characterized by a magnetic order  with propagation vector $(0,\pm 1/3,0)$ and consisting of six \ch{Mn}-\ch{Se} layers along the b-axis.
Its nonmagnetic phase is centrosymmetric with space group $\mathrm{Pa\overline{3}}$ (No.205).
We take the magnetic order of \ch{MnSe2} determined by neutron scattering experiments~\cite{1987_SSC_Chattopadhyay} as state~1, which belongs to magnetic space group $\mathrm{Pca'2_1'}$.
It hosts an intrinsic symmetry operator $\hat{r} = [-1||\hat{M}_{010}]$, which belongs to the multiferroic stabilizer group. 
This magnetic order breaks spatial inversion symmetry, rendering the system a type-II multiferroic with a magnetically induced polarization~\cite{2011_PRL_Xiang, 2012_PRL_Lu}.
Within our GTFC framework, the switching operator that connects two degenerate multiferroic states follows directly as $\hat{t} = \mathcal{\hat{T}\hat{P}} \cdot \hat{r} = [1||\hat 2_{010}]$. 
Applying $\hat{t}$ to state~1 predicts a symmetry related partner, state~2, which shares the same magnetic space group and is energetically degenerate with state~1. 
To identify the physical pathway for switching between these two states, we computationally examined two distinct mechanisms:
One path, which fixes the magnetic order and displaces the coordinated Se ions, yields a prohibitively high energy barrier ($\sim $1 eV) (details are provided in the Supplemental Material Fig.~S1~\cite{SM_file}).
The alternative path flips only the magnetic moments on the second and fifth \ch{Mn}-\ch{Se} layers while preserving the ionic framework, thereby avoiding lattice distortions and the associated large energy cost. 
This purely magnetic pathway yields a low barrier of \(1.31~meV/f.u.\), demonstrating the physical accessibility of the predicted switching mechanism.
Specifically, they show a maximum band splitting of 38.6~meV along the \(R\)--\(\Gamma\) direction [Fig.~\ref{Fig4}(c)] and opposing polarizations of \(\pm 0.104~\mu\mathrm{C}/\mathrm{cm}^2\) along the \textbf{b}-axis [Fig.~\ref{Fig4}(d)].
The low-energy switching of NRSS via magnetoelectric coupling in \ch{MnSe2} demonstrates the power of our symmetry based framework, which provides a general guide for designing control pathways in diverse collinear magnets.

\textit{Summary and discussion}——In summary, we have developed a general symmetry-based theory for achieving ferroelectric control of spin splitting in collinear antiferromagnets.
By decomposing the stabilizer group of multiferroic states into dimensionality-dependent and material-intrinsic components, we systematically enumerate all possible switching operators that simultaneously reverse polarization and spin texture.
Our work provides a unified framework for predicting the coupled reversal of polarization and NRSS, thereby extending the design space far beyond the widely studied 2D systems with out-of-plane polarization.
Guided by this strategy, we predict ferroelectric switching mechanisms in three unreported materials: (6,14)-ZGNRs,~\ch{Nb3I8}, and~\ch{MnSe2}, which are subsequently validated by our first-principles calculations.

In fact, the group-theoretical framework established here provides a versatile roadmap for discovering and designing ferroelectric switchable pathways for these collinear AFMs with NRSS beyond the demonstrated mechanisms. 
More broadly, it also offers a unified symmetry-based explanation for NRSS switching driven by other external operators, such as specific atomic displacements~\cite{2025_NL_Wang,2025_PRL_Gu} and electric field control~\cite{2025_NL_Wang_ding}.
It naturally guides the search for candidate materials through symmetry screening, and suggests unexplored switching pathways, such as those mediated by ionic displacements or symmetry-lowering structural distortions, in both low-dimensional and 3D systems, including those with nonpolar point groups where ferroelectricity is magnetically induced. 
Ultimately, this work lays a foundational design principle for the development of energy-efficient, multifunctional spintronic devices harnessing the interplay between ferroelectricity and collinear AFMs with NRSS.  

\textit{Acknowledgments}——We acknowledge financial support from the National Key R\&D Program of China (No. 2022YFA1402901), NSFC (grants No. 12188101), Shanghai Science and Technology Program (No. 23JC1400900), the Guangdong Major Project of the Basic and Applied Basic Research (Future functional materials under extreme conditions--2021B0301030005), Shanghai Pilot Program for Basic Research---FuDan University 21TQ1400100 (23TQ017), the robotic AI-Scientist platform of Chinese Academy of Science, and New Cornerstone Science Foundation.

\bibliography{reference}

\end{document}